\input{aipcheck}

\documentclass[
    ,final            
  ]
  {aipproc}

\layoutstyle{8x11single}

\def\jpsi{{J/\psi}}

\def\chico{{\chi_{c1}}}
\def\chict{{\chi_{c2}}}

\def\p0{{\bigl.^3\hspace{-1mm}P^{[8]}_0}}

\def\mosochij{{\langle\mathcal{O}^{\chi_{cJ}}(\bigl.^3\hspace{-1mm}S_1^{[8]})\rangle}}
\def\mosochiz{{\langle\mathcal{O}^{\chi_{c0}}(\bigl.^3\hspace{-1mm}S_1^{[8]})\rangle}}
\def\mopschij{{\langle\mathcal{O}^{\chi_{cJ}}(\bigl.^3\hspace{-1mm}P_J^{[1]})\rangle}}

\def\to{\rightarrow}

\def\bqa{\begin{eqnarray}}
\def\eqa{\end{eqnarray}}
\def\bc{\begin{center}}
\def\bc{\end{center}}
\def\HELACOnia{{\sc \small HELAC-Onia}}

\def\be{\begin{equation}}
\def\ee{\end{equation}}
\def\bea{\begin{eqnarray}}
\def\eea{\end{eqnarray}}

\def\th{\theta}

\begin{document}

\title{Probing Heavy Quarkonium Production Mechnism: $\chi_c$ polarization}


\author{Hua-Sheng Shao}{
  address={PH Department, TH Unit, CERN, CH-1211 Geneva 23, Switzerland}
}



\begin{abstract}
The necesscity of the color-octet mechanism in describing heavy quarkonium production is a longstanding puzzle. Compared to the yields of heavy quarkonium, its polarizations should be a sensitive observable to pin down the color-octet contributions. In this talk, I will focus on the $\chi_c$ polarization in hadroproduction processes, which may provide a unique test for the color-octet mechanism in nonrelativistic QCD.
\end{abstract}

\maketitle


\section{Introduction}

Since the first discovery of $J/\psi$ in 1974~\cite{Aubert:1974js,Augustin:1974xw}, heavy quarkonium physics has played an important role in revealing and investigating QCD at the interplay between the perturbative regime and non-perturbative regime. However, due to the complications, till now, it is still unable to pin down its production mechanism, especially the so-called color-octet (CO) mechanism in non-relativistic QCD (NRQCD)~\cite{Bodwin:1994jh}.

For a long time, the relatively satisfactory comparisons between the color-singlet theoretical postdictions/predictions and experimental data (like heavy quarkonium production at B factories~\cite{Zhang:2006ay,Gong:2009ng,Ma:2008gq,Gong:2009kp,He:2009uf,Jia:2009np,Zhang:2009ym,Shao:2014rwa}, in photoproduction~\cite{Kramer:1994zi,Baranov:2002cf,Kniehl:2006sk}, in fixed-target production~\cite{Petrelli:1997ge,Maltoni:2006yp} as well as at hadron colliders~\cite{Brodsky:2009cf,Lansberg:2010kh}) make people doubt on the importance of CO contributions. However, the measurements of $J/\psi$ and $\psi(2S)$ production at the Tevatron~\cite{Abe:1992ww,Abe:1997yz,Abe:1997jz,Abulencia:2007us}  and the LHC~\cite{Aad:2011sp,Chatrchyan:2011kc,Aaij:2011jh,LHCb:2012af,Aaij:2012ag,Chatrchyan:2013cla,Aaij:2013nlm,Abelev:2011md,Aaij:2014qea,Chatrchyan:2013cla} with large $p_T$ transfer indeed indicate the necessity of CO contributions~\cite{Artoisenet:2007xi,Gong:2008ft,Gong:2008hk,Gong:2008zz,Butenschoen:2009zy,Ma:2010vd,Ma:2010yw,Butenschoen:2010rq,Ma:2010jj,Butenschoen:2011ks,Butenschoen:2011yh,Butenschoen:2012px,Chao:2012iv,Gong:2012ug,Shao:2014fca,Bodwin:2014gia,Faccioli:2014cqa,Gong:2008zz,Lansberg:2011hi}, in which the {\it smoking gun} signature is the polarization measurements. The unpolarized pattern of $J/\psi$ and $\psi(2S)$ was though to be a challenge to NRQCD for a long time. On contrast to S-wave charmoniua, P-wave charmonia are usually overlooked. In this talk, I will focus on the polarizations of $\chi_{c1}$ and $\chi_{c2}$. From the lessons of $\jpsi$ and $\psi(2S)$ production, one should also expect that the polarizations of $\chi_{c1}$ and $\chi_{c2}$ may provide an unique test on the CO mechanism (COM).

Compared to $\jpsi$ and $\psi(2S)$ production, there are several motivations to consider $\chi_c$ production as a complementary study. First, unlike the case of $\jpsi$, there is no significant feed-down contributions to prompt $\chi_c$ production. It makes the analysis simpler. Second, only one independent leading CO LDME $\mosochij$ should be determined from the experimental data. In other words, one can expect more predictive power of $\chi_c$ than of $\jpsi$ and $\psi(2S)$--which need to know three leading CO LDMEs. On the theoretical side, the unresolved problem of infrared (IR) divergences for P-wave states in color-singlet model (CSM) can be naturally overcomed by absorbing IR poles into the renormalization group evolution of CO S-wave states. From this point of view, COM is necessary for understanding $\chi_c$ production. Moreover, it is also necessary to investigate $\chi_c$ polarization, since almost $30\%$ of prompt $\jpsi$ is from the feed-down contribution via $\chi_c\to\jpsi+\gamma$.

In this talk, I will first generalize the spin-entangled decay amplitudes $\chi_c\to\jpsi+\gamma$ by including the impact of higher order multipole transitions. It will be quite useful for Monte Carlo simulations to implement in generators, like \HELACOnia~\cite{Shao:2012iz}. Before going into any phenomenological analysis, we will first fix the values of CO LDMEs from the yields data of $\chi_c$ at the Tevatron. Finally, we present the predictions for the polarizations of $\chi_{c1}$ and $\chi_{c2}$ production at the LHC, and propose the experimentlists to measure these observables.

\section{Spin-entangled decay of $\chi_c\to\jpsi+\gamma$}

In Ref.~\cite{Shao:2012fs}, we presented an extensive discussion of the angular distributions of $\chi_c$ decay. From the effective decay vertices, we derived the general formula for $\chi_c \to \jpsi+\gamma$ and the subsequence decay $\chi_c \to \jpsi+\gamma \to \ell^+\ell^-+\gamma$. We presented the final angular distributions with the higher-order multipole transition contributions in the appendices. However, it might be not sufficient for Monte Carlo simulations, which require the full knowledge of the helicity amplitudes with the (non-negligible) higher-order mulitpole transitions. We will establish the complete set of the helicity amplitudes for $\chi_c \to \jpsi+\gamma$ in this section.

Following the notations and conventions in Ref.~\cite{Shao:2012fs}, we can derive the following helicity amplitudes for $\chi_{c1}\to \jpsi+\gamma$:

\bqa
\mathcal{M}_{+++}=\mathcal{M}_{---}^*=\left(a^{J=1}_1+a^{J=1}_2\right)\frac{e^{-i\phi}\sin\th}{2\sqrt{2}},\nonumber\\
\mathcal{M}_{+--}=\mathcal{M}_{-++}^*=-\left(a^{J=1}_1+a^{J=1}_2\right)\frac{e^{3i\phi}\sin\th}{2\sqrt{2}},\nonumber\\
\mathcal{M}_{+0+}=-\mathcal{M}_{-0-}=-\left(a^{J=1}_1-a^{J=1}_2\right)\frac{\sin^2{\frac{\th}{2}}}{2},\nonumber\\
\mathcal{M}_{+0-}=-\mathcal{M}_{-0+}^*=\left(a^{J=1}_1-a^{J=1}_2\right)\frac{e^{2i\phi}\cos^2{\frac{\th}{2}}}{2},\nonumber\\
\mathcal{M}_{0++}=-\mathcal{M}_{0--}^*=-\left(a^{J=1}_1+a^{J=1}_2\right)\frac{e^{-2i\phi}\cos\th}{2},\nonumber\\
\mathcal{M}_{00+}=\mathcal{M}_{00-}^*=\left(a^{J=1}_1-a^{J=1}_2\right)\frac{e^{-i\phi}\sin\th}{2\sqrt{2}},
\eqa
where $\th,\phi$ are the polar and azimuthal angles of the decay product photon $\gamma$ in the rest frame of $\chi_c$. Symbols $a^{J=1}_1$ and $a^{J=1}_2$ denote the electric dipole (E1) transition amplitude and the magnetic quadrupole (M2) transition amplitude for $\chi_{c1}\to\jpsi+\gamma$.These amplitudes are assumed to be real and to be normalized $\left(a^{J=1}_1\right)^2+\left(a^{J=1}_2\right)^2=1$. In the above equation, we do not apply the assumption of $m_{\chi_{c1}}\simeq m_{\jpsi}$, and it is rigorous.

Along the same line, for $\chi_{c2}\to\jpsi+\gamma$, we have
\bqa
\mathcal{M}_{2++}=\mathcal{M}_{-2--}=\sqrt{\delta_0}\frac{\sqrt{3}\sin^2{\th}}{4},\nonumber\\
\mathcal{M}_{2+-}=\mathcal{M}_{-2-+}^*=\sqrt{1-\delta_0-\delta_1}\frac{e^{2i\phi}\cos^4{\frac{\th}{2}}}{\sqrt{2}},\nonumber\\
\mathcal{M}_{2--}=\mathcal{M}_{-2++}^*=\sqrt{\delta_0}\frac{\sqrt{3}e^{4i\phi}\sin^2{\th}}{4},\nonumber\\
\mathcal{M}_{2-+}=\mathcal{M}_{-2+-}^*=\sqrt{1-\delta_0-\delta_1}\frac{e^{2i\phi}\sin^4{\frac{\th}{2}}}{\sqrt{2}},\nonumber\\
\mathcal{M}_{20+}=-\mathcal{M}_{-20-}^*=-\sqrt{\delta_1}\sqrt{2}e^{i\phi}\cos{\frac{\th}{2}}\sin^3{\frac{\th}{2}},\nonumber\\
\mathcal{M}_{20-}=-\mathcal{M}_{-20+}^*=-\sqrt{\delta_1}\sqrt{2}e^{3i\phi}\cos^3{\frac{\th}{2}}\sin{\frac{\th}{2}},\nonumber\\
\mathcal{M}_{1++}=-\mathcal{M}_{-1--}^*=-\sqrt{\delta_0}\frac{\sqrt{3}e^{-i\phi}\sin{2\th}}{4},\nonumber\\
\mathcal{M}_{1+-}=-\mathcal{M}_{-1-+}^*=\sqrt{1-\delta_0-\delta_1}\sqrt{2}e^{i\phi}\cos^3{\frac{\th}{2}}\sin{\frac{\th}{2}},\nonumber\\
\mathcal{M}_{1--}=-\mathcal{M}_{-1++}^*=-\sqrt{\delta_0}\frac{\sqrt{3}e^{3i\phi}\sin{2\th}}{4},\nonumber\\
\mathcal{M}_{1-+}=-\mathcal{M}_{-1+-}^*=-\sqrt{1-\delta_0-\delta_1}\sqrt{2}e^{i\phi}\cos{\frac{\th}{2}}\sin^3{\frac{\th}{2}},\nonumber\\
\mathcal{M}_{10+}=\mathcal{M}_{-10-}=\sqrt{\delta_1}\frac{\left(1+2\cos{\th}\right)\sin^2{\frac{\th}{2}}}{\sqrt{2}},\nonumber\\
\mathcal{M}_{10-}=\mathcal{M}_{-10+}^*=\sqrt{\delta_1}\frac{e^{2i\phi}\cos^2{\frac{\th}{2}}\left(2\cos{\th}-1\right)}{\sqrt{2}},\nonumber\\
\mathcal{M}_{0++}=\mathcal{M}_{0--}^*=\sqrt{\delta_0}\frac{e^{-2i\phi}(1+3\cos{2\th})}{4\sqrt{2}},\nonumber\\
\mathcal{M}_{0+-}=\mathcal{M}_{0-+}=\sqrt{1-\delta_0-\delta_1}\frac{\sqrt{3}\sin^2{\th}}{4},\nonumber\\
\mathcal{M}_{00+}=-\mathcal{M}_{00-}^*=-\sqrt{\delta_1}\frac{\sqrt{3}e^{-i\phi}\sin{2\th}}{4},
\eqa
where the coefficents $\delta_0$ and $\delta_1$ can be expressed as the polynomials of E1, M2 and electric octupole (E3) transition amplitudes $a^{J=2}_1,a^{J=2}_2,a^{J=2}_3$,\footnote{We assume the multipole transition amplitudes are real and normalized $\left(a^{J=2}_1\right)^2+\left(a^{J=2}_2\right)^2+\left(a^{J=2}_3\right)^2=1$.}

\bqa
\delta_0\equiv \frac{1+2a^{J=2}_1\left(\sqrt{5}a_2^{J=2}+2a_3^{J=2}\right)+4a^{J=2}_2\left(a^{J=2}_2+\sqrt{5}a^{J=2}_3\right)
+3\left(a^{J=2}_3\right)^2}{10},\nonumber\\
\delta_1\equiv \frac{9+6a^{J=2}_1\left(\sqrt{5}a^{J=2}_2-4a^{J=2}_3\right)-4a^{J=2}_2\left(a^{J=2}_2
+2\sqrt{5}a^{J=2}_3\right)+7\left(a^{J=2}_3\right)^2}{30}
\eqa
With Eqs.(3,7,11) in Ref.~\cite{Shao:2012fs}, the above equations render the angular distributions of $\chi_{c1}\to \jpsi+\gamma$ and $\chi_{c2}\to \jpsi+\gamma$ established in appendices A and B of Ref.~\cite{Shao:2012fs}. Combining the well-known vector currents form helicity amplitudes for $\jpsi\to\ell^+\ell^-$, one is easily able to derive the angular distributions for the cascade decays $\chi_c\to\jpsi+\gamma\to \ell^+\ell^-+\gamma$, which were established in appendix C of Ref.~\cite{Shao:2012fs}. In the following, we only consider the polar angle distributions. For the $\jpsi$ or $\gamma$ angular distributions, we have
\bqa
\frac{d\mathcal{N}^{\chi_{cJ}}}{d\cos\th} \sim 1+\sum_{k=1}^{J}{\lambda_{k\th}\cos^{2k}{\th}}.
\label{eq:th}
\eqa
We will refrain the expressions for the polar asymmetry coefficeints $\lambda_{k\th}$ in terms of $\chi_c$ production spin density matrix elements and the multipole amplitudes, but refering the interested readers to the corresponding formula presented in
Refs.~\cite{Shao:2012fs,Shao:2014fca}. Similarly, the lepton polar angle $\th^{\prime}$ dependence can be sketched as
\bqa
\frac{d\mathcal{N}^{\chi_{cJ}}}{d\cos\th^{\prime}} \sim 1+\lambda_{\th^{\prime}}\cos^2{\th^{\prime}}.
\label{eq:thp}
\eqa
One can also find the corresponding formula in Refs.~\cite{Shao:2012fs,Shao:2014fca} for the coefficient $\lambda_{\th^{\prime}}$.

The multipole amplitudes have been measured by CLEO~\cite{Artuso:2009aa}, Crystal Ball~\cite{Oreglia:1981fx}, E760~\cite{Armstrong:1993fk}, E835~\cite{Ambrogiani:2001jw} Collaborations. Unfortunately, the measured values for the higher-order multipole amplitudes are still inconsistent among the various measurements. In the following, we will take the values fitted by CLEO~\cite{Artuso:2009aa} Collaboration as our input values. In sepcific, we have
\bqa
a^{J=1}_2&=&-6.26\times 10^{-2},\nonumber\\
a^{J=2}_2&=&-9.3\times 10^{-2},\nonumber\\
a^{J=2}_3&=&0,
\eqa
where we have taken the E3 amplitude for $\chi_{c2}$ decay to be zero from the single quark radiation hypothesis~\cite{Karl:1980wm,Olsson:1984zm}. The nonvanishing of these higher-order multipole amplitudes are important for determining the $\jpsi$ or $\gamma$ angular distributions, while it only mildly changes the lepton angular distributions in the cascade decays $\chi_c\to\jpsi+\gamma\to \ell^+\ell^-+\gamma$.

\section{CO LDMEs}

The values of CO LDMEs $\mosochij$ can be determined by fitting the Tevatron data of $\sigma(\chi_{c2})/\sigma(\chi_{c1})$~\cite{Abulencia:2007bra} after applying the spin symmetry relation $\mosochij=(2J+1)\mosochiz$. At NLO in $\alpha_s$, it was first extracted in Ref.~\cite{Ma:2010vd}. We estimate the CS LDMEs $\mopschij$ via potential model~\cite{Eichten:1995ch} as $\mopschij=(2J+1)3|R^{\prime}(0)|^2/4\pi$ and $|R^{\prime}(0)|^2=0.075~{\rm GeV}^5$. The CO LDMEs can be determined as $\mosochij=(2J+1)\times (2.2^{+0.48}_{-0.32})\times 10^{-3}~{\rm GeV}^3$~\cite{Ma:2010vd}. Several possible comparisons for $\chi_c$ yields to the LHC data have been done in Refs.~\cite{Shao:2014fca,LHCb:2012ac,LHCb:2012af,Chatrchyan:2012ub} with these non-perturbative LDMEs. We found that good agreements between theory and experiment were achieved.

\section{Polarizations}

\begin{figure}[ht]
\begin{minipage}[c]{0.25\textwidth}
\begin{center}
\includegraphics[width=\textwidth]{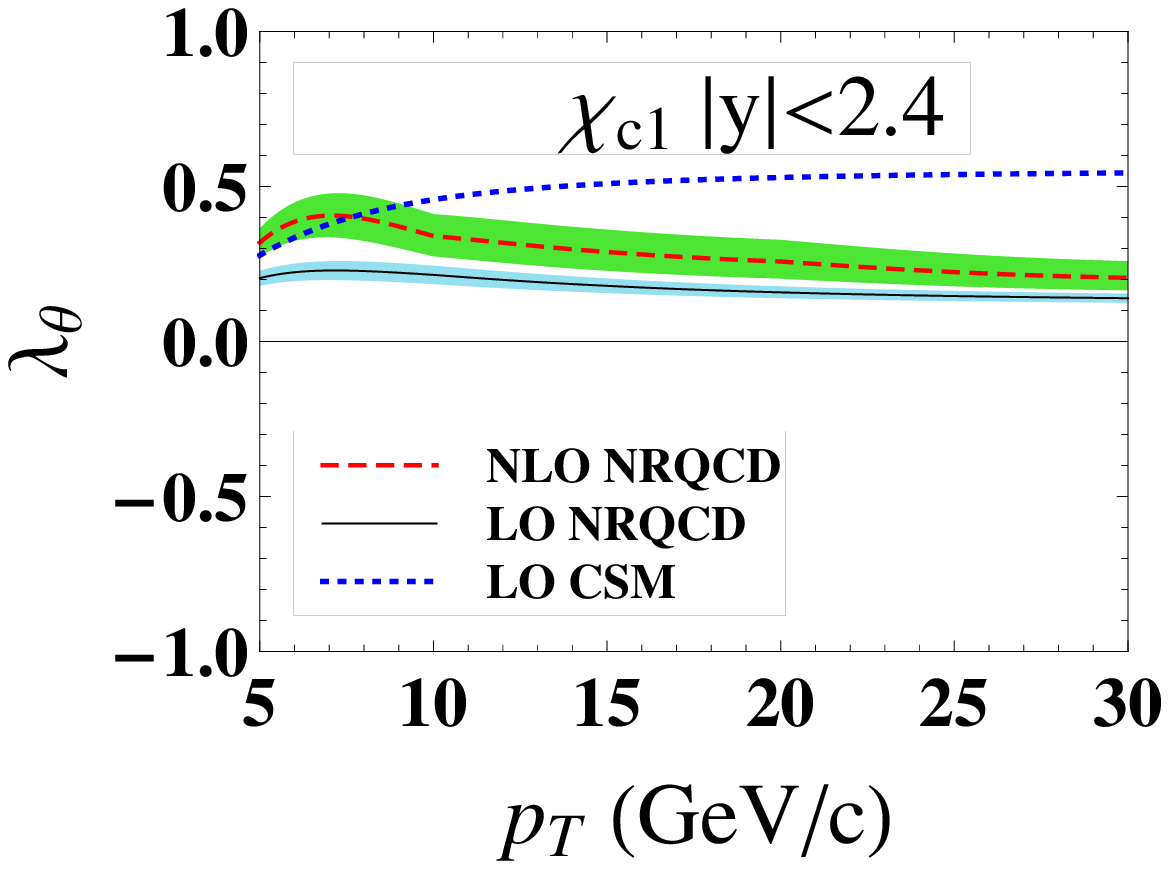}
(a)
\end{center}
\end{minipage}
\begin{minipage}[c]{0.25\textwidth}
\begin{center}
\includegraphics[width=\textwidth]{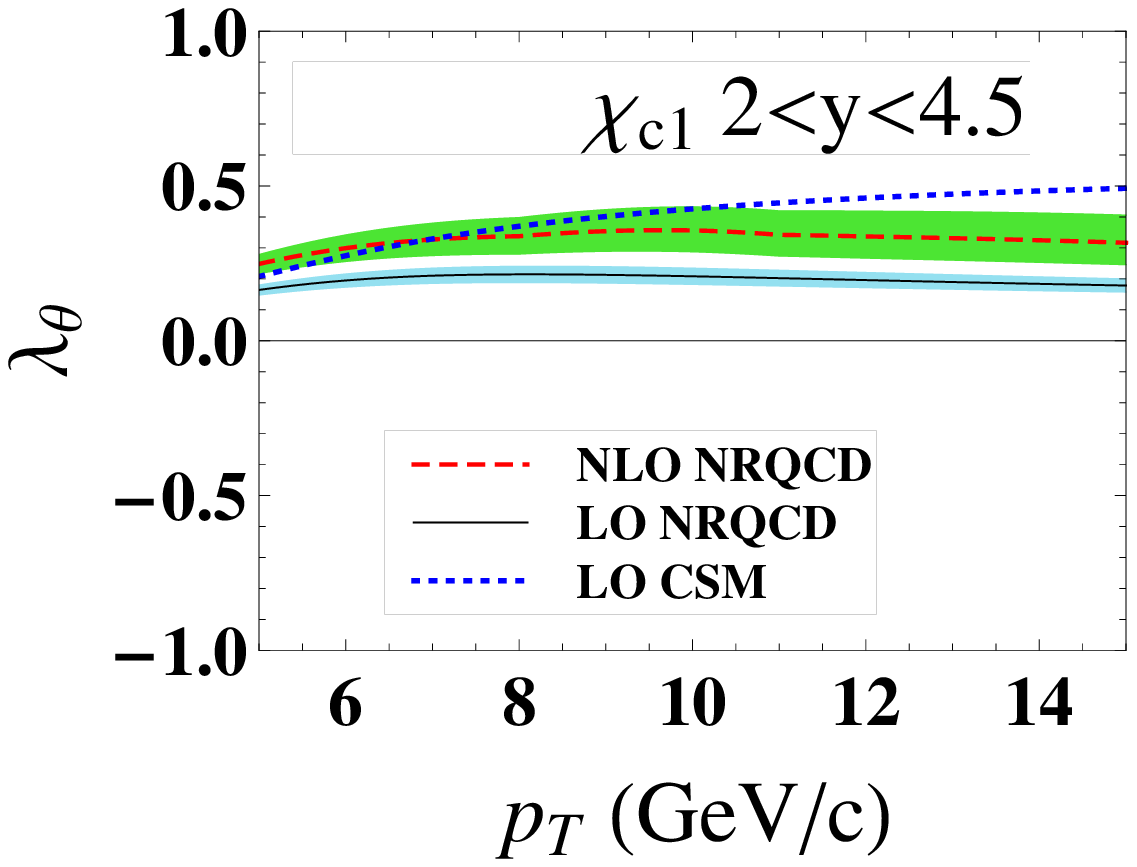}
(b)
\end{center}
\end{minipage}
\begin{minipage}[c]{0.25\textwidth}
\begin{center}
\includegraphics[width=\textwidth]{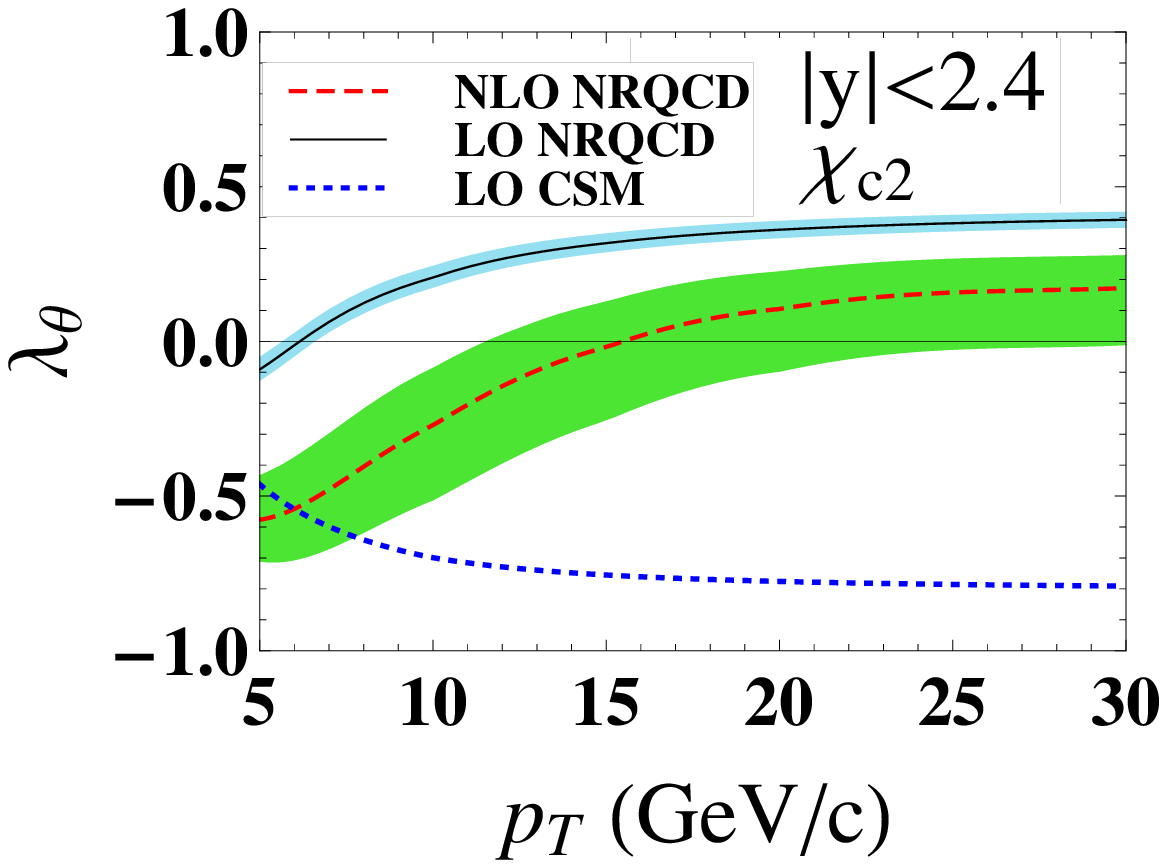}
(c)
\end{center}
\end{minipage}
\begin{minipage}[c]{0.25\textwidth}
\begin{center}
\includegraphics[width=\textwidth]{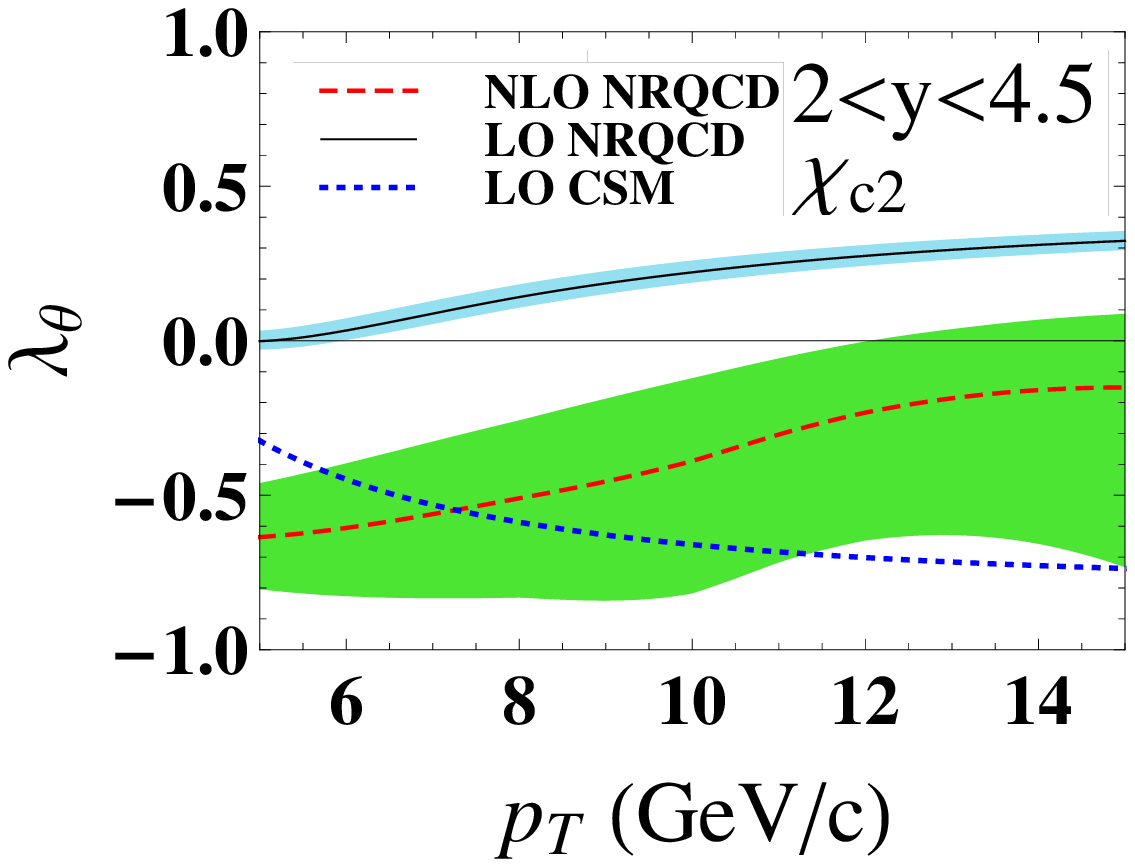}
(d)
\end{center}
\end{minipage}
\caption{\label{fig:thpol} The $p_{T}$ spectra of
$\lambda_{\th}$ for $\jpsi$ or photon angular distributions from
$\chico\to\jpsi\gamma$ (a-b) and
$\chict\to\jpsi\gamma$ (c-d) in the helicity
frame at the LHC with $\sqrt{S}=7~\rm{TeV}$.  The plots are taken from Ref.~\cite{Shao:2014fca} with kind permission. Copyright 2014 American Physics Society.}
\end{figure}

In Ref.~\cite{Shao:2014fca}, we managed to present a first rigorous theoretical prediction for the $\chi_c$ polarization observables at the LHC. We displayed the $\chi_c$ $p_T$ distributions for the $\jpsi$ or photon polar asymmetry coefficients $\lambda_{\th}$ in Fig.\ref{fig:thpol}. I want to remind the reader that this observable is sensitive to the values of higher-order multipole amplitudes. Hence, the measurement of such observable at the LHC may cross check the measured values of the higher-order multipole amplitudes. For comparison, the LO NRQCD results and the LO CSM results are also included. At LO in $\alpha_s$, CO contribution is dominant. Hence, the LO NRQCD results can be viewed as the polarization behaviour determined by CO contribution. At NLO in $\alpha_s$, the CS component will partly cancel the CO component. In Fig.\ref{fig:thpol}, one can clearly see that the LO NRQCD results indeed share the different behaviours with the CSM result. Therefore, it is understood that the NLO NRQCD curves lie between LO NRQCD and LO CSM curves when $p_T>10~\rm{GeV}$. It may provide a good discrimination to determine the fraction of CO contributions in $\chi_c$ production at the LHC. In E1 approximation, the coefficeint $\lambda_{2\th}$  for $\chi_{c2}$ polarization in Eq.\ref{eq:th} is zero. Hence, we refrain ourselves to present a prediction for $\lambda_{2\th}$ here, because in anycase it is quite close to zero. A reweighting method proposed in Ref.~\cite{Shao:2012fs} may help to determine this observable on the experimental side.


\begin{figure}[ht]
\begin{minipage}[c]{0.25\textwidth}
\begin{center}
\includegraphics[width=\textwidth]{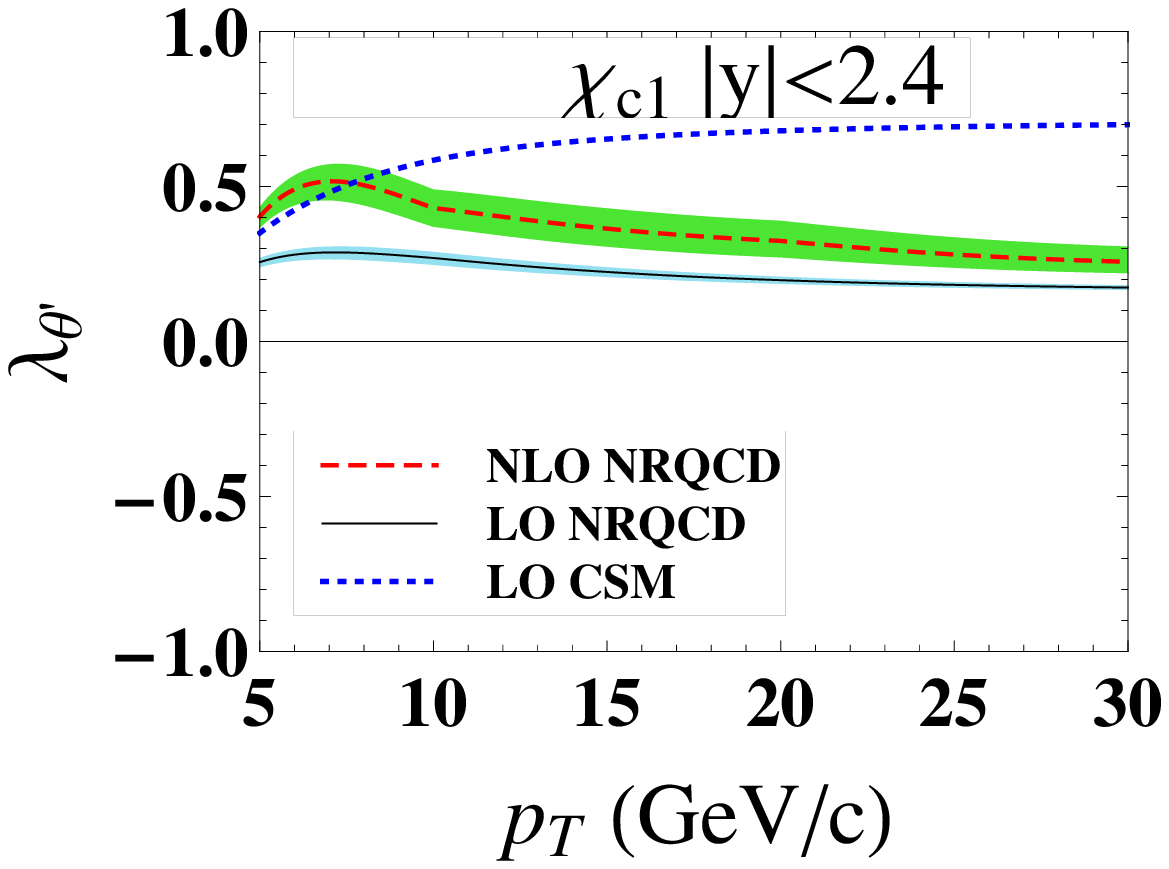}
(a)
\end{center}
\end{minipage}
\begin{minipage}[c]{0.25\textwidth}
\begin{center}
\includegraphics[width=\textwidth]{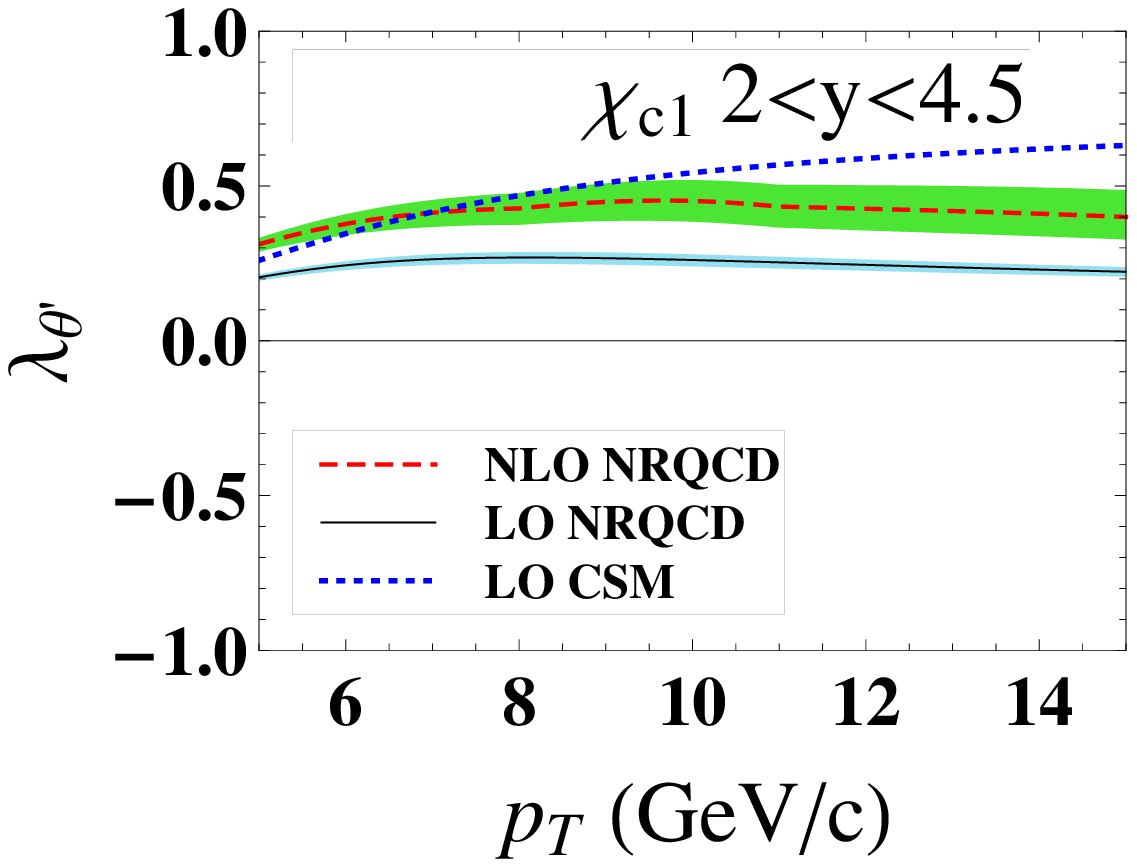}
(b)
\end{center}
\end{minipage}
\begin{minipage}[c]{0.25\textwidth}
\begin{center}
\includegraphics[width=\textwidth]{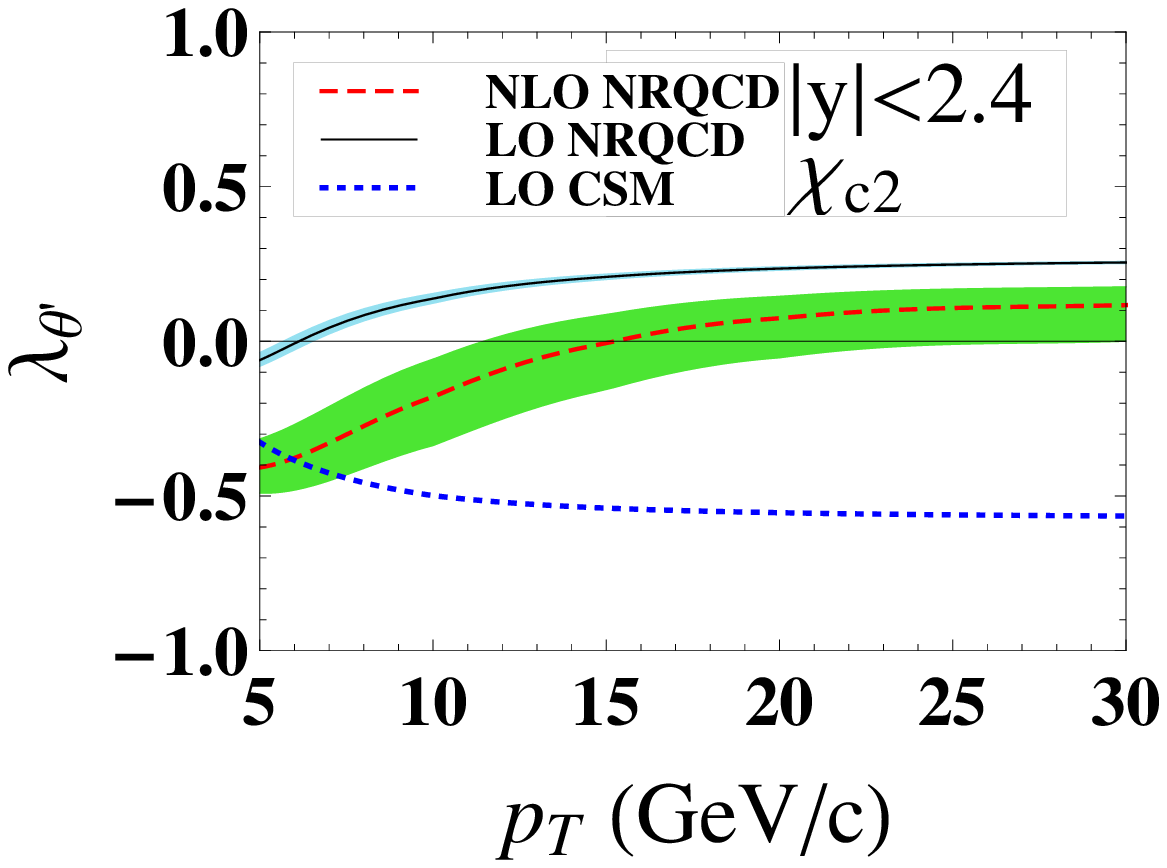}
(c)
\end{center}
\end{minipage}
\begin{minipage}[c]{0.25\textwidth}
\begin{center}
\includegraphics[width=\textwidth]{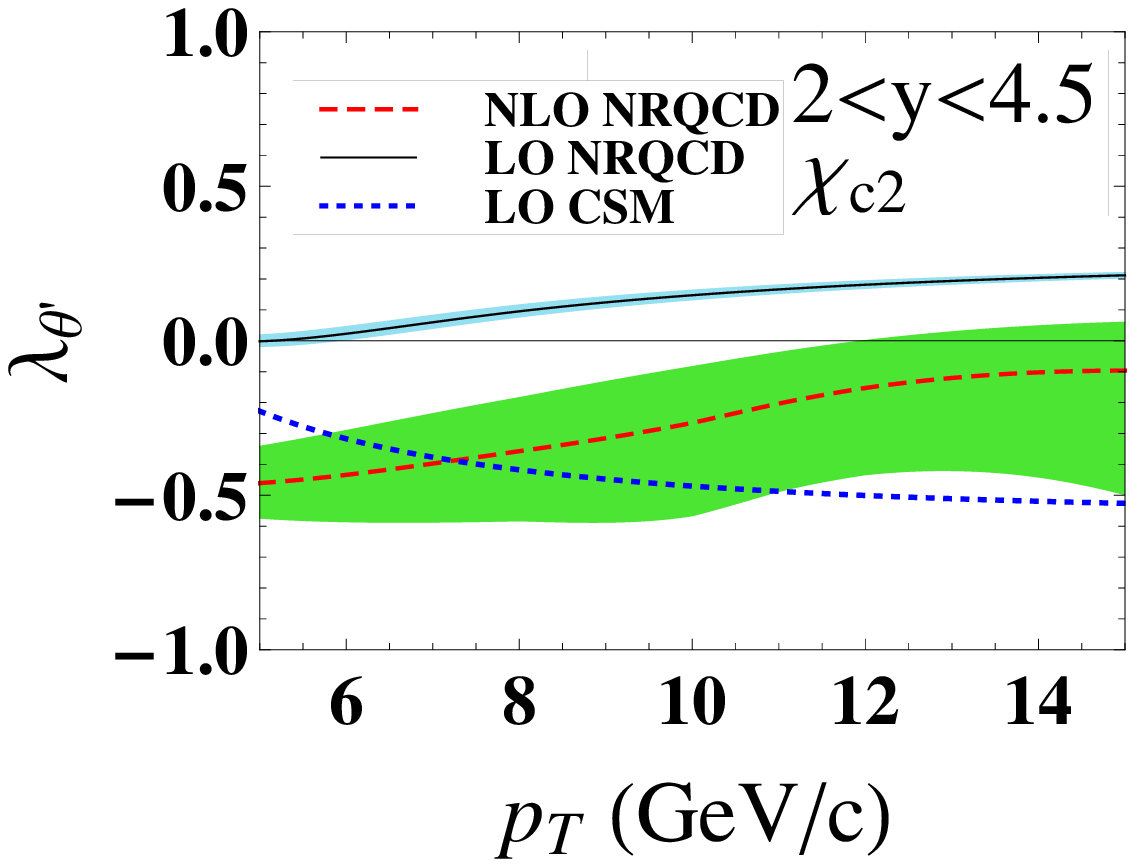}
(c)
\end{center}
\end{minipage}
\caption{\label{fig:thppol} The $p_{T}$ spectra of
$\lambda_{\th^{\prime}}$ for  dilepton angular
distributions from $\chico\to\jpsi\gamma\to
l^+l^-\gamma$ (a-b) and $\chict\to\jpsi\gamma\to
l^+l^-\gamma$ (c-d) in the helicity frame at
the LHC with $\sqrt{S}=7~\rm{TeV}$. The plots are taken from Ref.~\cite{Shao:2014fca} with kind permission. Copyright 2014 American Physics Society.}
\end{figure}

For the coefficient $\lambda_{\th^{\prime}}$ in the dilepton angular distribution Eq.\ref{eq:thp}, the $p_T$ spectra are shown in Fig.\ref{fig:thppol}. Unlike $\lambda_{\th}$, $\lambda_{\th^{\prime}}$ should be insensitive to the values of the higher-order multipole amplitudes. Hence, it can be compared to the experimental data without receiving any significant uncertainty from the inconsistent measurements of $a^{J=1}_2,a^{J=2}_2,a^{J=2}_3$. Similar to $\lambda_{\th}$, the behaviour of CO is different to that of CS in  $\lambda_{\th^{\prime}}$, and it provides another good observable to distuinguish COM and CSM. Finally, I would like to emphasize that it is the same observable to show the feed-down part of the $\jpsi$ polarization from $\chi_c \to \jpsi+\gamma$.

\section{Conclusion}

In this talk, I mainly focus on the $\chi_c$ polarization, which might provide an unique test to COM at the LHC. I first derived a set of helicity amplitudes for the spin-entangled decay of $\chi_c\to \jpsi+\gamma$, which is completely new and useful to implement into generators for Monte Carlo simulations. Then, I presented our theoretical predictions for the polarizations of $\chi_{c1}$ and $\chi_{c2}$ production at the LHC in the NRQCD framework at NLO in $\alpha_s$. These polarization observables may be important to determe/test the CO contributions in $\chi_c$ hadroproduction. Moreover, the $\jpsi$ or photon angular distributions may also provide a possible way to extract the higher-order mutlipole amplitudes at the LHC, which is still poorly known.


\begin{theacknowledgments}
The work is supported by the National Natural Science Foundation of China (No 11075002, No 11021092)  and ERC grant 291377 ``LHCtheory: Theoretical predictions and analyses of LHC physics: advancing the precision frontier". 
\end{theacknowledgments}






\IfFileExists{\jobname.bbl}{}
 {\typeout{}
  \typeout{******************************************}
  \typeout{** Please run "bibtex \jobname" to optain}
  \typeout{** the bibliography and then re-run LaTeX}
  \typeout{** twice to fix the references!}
  \typeout{******************************************}
  \typeout{}
 }

\end{document}